# Hardware Architecture Design of Model-Based Image Reconstruction Towards Palm-size Photoacoustic Tomography

Yuwei Zheng, Zijian Gao, Yuting Shen, Jiadong Zhang, Daohuai Jiang, Fengyu Liu, Feng Gao*, and Fei Gao*, *Member, IEEE*

***Abstract*—Photoacoustic (PA) imaging technology combines the advantages of optical imaging and ultrasound imaging, showing great potential in biomedical applications. Many preclinical studies and clinical applications urgently require fast, high-quality, low-cost and portable imaging system. Translating advanced image reconstruction algorithms into hardware implementations is highly desired. However, existing iterative PA image reconstructions, although exhibit higher accuracy than delay-and-sum algorithm, suffer from high computational cost. In this paper, we introduce a model-based hardware acceleration architecture based on superposed Wave (s-Wave) for palm-size PA tomography (palm-PAT), aiming at enhancing both the speed and performance of image reconstruction at a much lower system cost. To achieve this, we propose an innovative data reuse method that significantly reduces hardware storage resource consumption. We conducted experiments by FPGA implementation of the algorithm, using both phantoms and in vivo human finger data to verify the feasibility of the proposed method. The results demonstrate that our proposed architecture can substantially reduce system cost while maintaining high imaging performance. The hardware-accelerated implementation of the model-based algorithm achieves a speedup of up to approximately 270 times compared to the CPU, while the corresponding energy efficiency ratio is improved by more than 2700 times.**

## I. INTRODUCTION

PHOTOACOUSTIC (PA) imaging (PAI) is an emerging hybrid medical imaging modality that leverages the photoacoustic effect, where optical absorption by biological tissues induces thermoelastic expansion, subsequently generating ultrasound waves [1], [2]. This technique uniquely combines the high optical contrast with the deep tissue penetration capabilities of ultrasound [3]-[5] offering promising applications in both pre-clinical research and clinical settings. The increasing demand for portable devices in outdoor or emergency scenarios necessitates rapid and high-quality imaging solutions, posing stringent requirements on the cost, miniaturization, and performance of PA imaging systems.

In PA image reconstruction, the delay-and-sum (DAS) beamforming algorithm is widely adopted due to its simplicity and fast imaging capabilities. However, DAS is prone to severe artifacts, especially under limited-view or sparse sampling conditions. To address these limitations, several variants of DAS, such as delay-multiply-and-sum (DMAS) [6], [7] and delay-and-sum with coherence factor (DAS-CF) [8], have been developed and accelerated using field-programmable gate arrays (FPGAs), thereby enhancing reconstruction speed [9]. Another commonly used one step method is time-reversal reconstruction [10], which, despite its efficiency, still suffers from critical artifacts.

Currently, FPGA-accelerated reconstruction and signal processing techniques predominantly focus on simple DAS-based methods [9]-[12] and back projection (BP) algorithm [13]. While these methods offer computational efficiency, they often fall short of delivering high image quality. In contrast, iterative model-based reconstruction algorithms [14] have demonstrated superior image quality by solving complex numerical equations, albeit at a significant computational cost. Although Graphics Processing Units (GPUs) [15] can accelerate these iterative methods [16], [17], including compressed sensing [18] and deep learning frameworks [19]-[21], they do not satisfy the low-cost, low-power consumption, and miniaturization requirements essential for portable PA imaging devices. Therefore, there is a pressing need to accelerate these iterative algorithms cost-effectively.

In this paper, we present a novel hardware architecture based on a model-based iterative algorithm [22], [23] that employs the superposed wave (s-Wave) technique [24], [25] to enhance PA image reconstruction while maintaining high image quality. Our approach introduces a novel data reuse method leveraging the geometric symmetry of ultrasound transducer distributions, significantly reducing on-chip storage requirements. The proposed design utilizes DAS as the backward model to

Yuwei Zheng, Yuting Shen, Fengyu Liu, Feng Gao and Fei Gao are with the School of Information Science and Technology, ShanghaiTech University, Shanghai, 201210 China. (e-mail: zhengyw2022@shanghaitech.edu.cn; shenyt@shanghaitech.edu.cn; liufy@shanghaitech.edu.cn; gaofeng@shanghaitech.edu.cn; gaofei@shanghaitech.edu.cn).

Zijian Gao was with the School of Information Science and Technology, ShanghaiTech University, Shanghai, 201210 China. He is now with the Faculty of Medicine, Chinese University of Hong Kong, Hong Kong, China. (e-mail: gaozj@shanghaitech.edu.cn).

Jiadong Zhang was with the School of Information Science and Technology, ShanghaiTech University, Shanghai, 201210 China. He is now with the Department of Biomedical Engineering, City University of Hong Kong, Hong Kong, China. (e-mail: zhangjd1@shanghaitech.edu.cn).

Daohuai Jiang was with the School of Information Science and Technology, ShanghaiTech University, Shanghai, 201210 China. He is now with the College of Photonic and Electronic Engineering, Fujian Normal University, Fuzhou, China. (e-mail: jiangdh@fjnu.edu.cn).

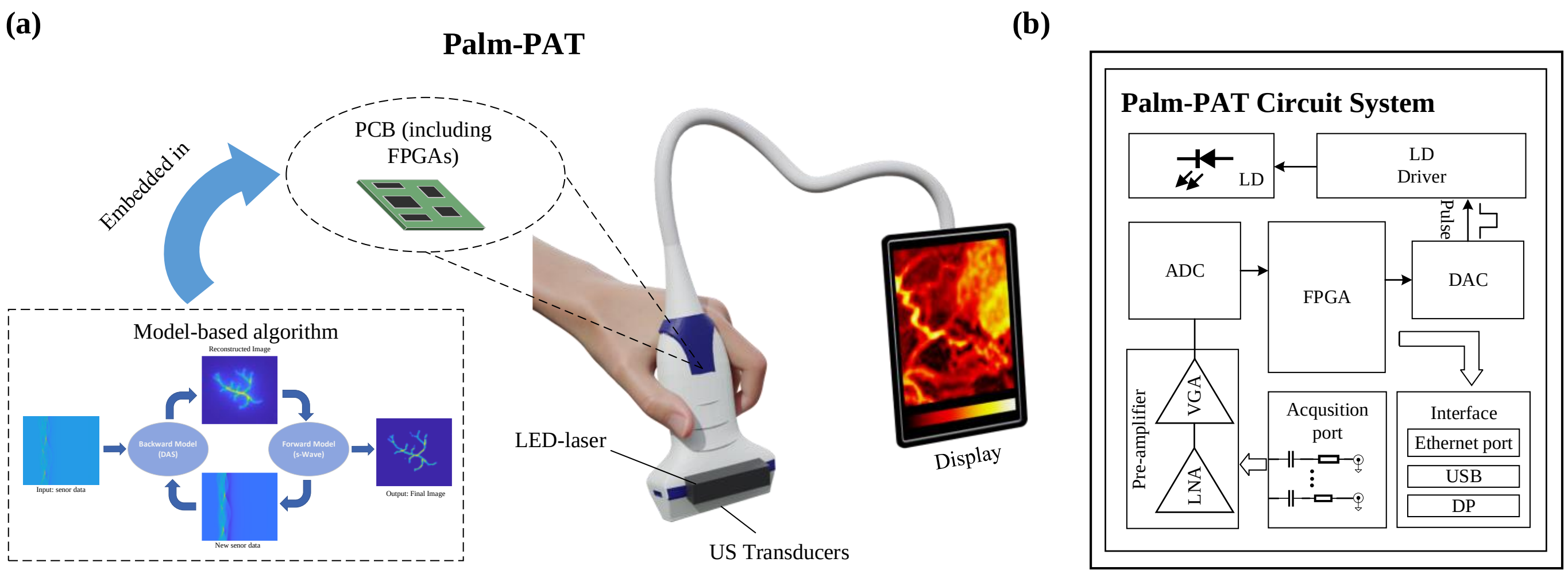

Fig. 1. An envisioned Palm-PAT system. It integrates the ultrasound transducer, LED laser, data acquisition module, image reconstruction hardware module, and high-speed data interfaces. (a) Diagram of the palm-PAT. PCB: printed circuit board, LED: light emitting diode, US Transducers: Ultrasound Transducers. (b) Palm-PAT circuit system. LD: laser diode, ADC: analog to digital converter, DAC: digital to analog converter, LNA: low noise amplifier, VGA: variable gain amplifier, USB: universal serial bus, DP: DisplayPort.

transform signals into images, while the s-Wave serves as the forward model to convert images back into signals, simplifying the traditional process of solving complex numerical equations into basic multiplication and summation operations, thereby reducing algorithmic complexity. These operations are implemented using multi-channel parallelism and a pipelining scheme to accelerate data processing. We conducted PA imaging of phantom and in vivo experiments on FPGA platforms to demonstrate the feasibility and potential applications of our proposed approach, indicating its promise for integration into palm-PAT system.

## II. Methods

### A. Overview of palm-PAT

Our envisioned palm-PAT system, as shown in Fig. 1(a), is designed to be entirely handheld. The front end of the system features ultrasound transducers for receiving PA signals, with an adjacent LED [26], [27] laser, which is able to operate at high repetition rates of several kHz [28] emitting pulsed laser light. The entire PA imaging hardware is housed within the device body and integrated onto printed circuit boards (PCB). A noteworthy aspect is the implementation of various PA image reconstruction algorithms (e.g., DAS, DMAS, Model-based) on FPGA (or application-specific integrated circuits (ASIC) chip in future work) which is integrated into the hardware system, enabling low-latency and real-time image reconstruction. The system framework of the palm-PAT is illustrated in Fig. 1(b). The front end includes a preamplifier, filter, and ADC for signal amplification, filtering, and analog-to-digital conversion. The FPGA is responsible for generating the laser diode drive signal, controlling the entire data acquisition process, performing image reconstruction, and managing data transmission. The interface section includes ports such as DP, USB, and ethernet and so on, facilitating connections to external devices like displays and personal computers (PCs) for real-time image display and high-speed data transmission.

### B. Algorithm Design

#### 1) Backward Model: Delay-and-Sum Algorithm

Delay-and-Sum (DAS) is the most commonly used method [9] for rapid PA image reconstruction. The specific description is shown in Algorithm 1.

**Algorithm 1** Delay-and-Sum

**Require:** Input sensor data $S$; the number of sensors N; ROI size $n \times m$; pixel position $(x_p, y_p)$; sensors position $(x_s, y_s)$; Sound speed c; the sample frequency $f_s$.

**Ensure:** Reconstructed image $S_{DAS}$.

$I_{DAS} = 0$
**for** $i \leftarrow 1, N$ **do**
  **for** $j \leftarrow 1, n \times m$ **do**
    $d_{ij} = \sqrt{(x_{pj} - x_{si})^2 + (y_{pj} - y_{si})^2}$
    $delay\,(x_{pj}, y_{pj}, i) = \frac{d_{ij}}{c} \cdot f_s$
    $I_{DAS}(j) = I_{DAS}(j) + S\,(i, delay\,(x_{pj}, y_{pj}, i))$
  **end for**
**end for**
**return** Reconstructed image $I_{DAS}$

It processes the input sensor data based on the spatial relationship between each ultrasound transducer element (sensor) and each pixel position. Before performing the reconstruction, the positions of the sensors are fixed so that we can get the distances between the sensors and various positions within the imaging region in advance. Therefore, the delay in DAS can be expressed as

$$delay\,(x,y,i) = \frac{d\,(x,y,i)}{c} \cdot f_s \tag{1}$$

where $d\,(x,y,i)$ represents the distance from position $(x,y)$ to the $i^{th}$ sensor. $c$ is the sound speed (1500 m/s), and $f_s$ is the sampling frequency. By extracting and summing the collected sensor data based on these delay values, we can obtain the final image values $I_{DAS}(x,y)$, which can be expressed as

$$I_{DAS}(x,y)=\sum_{i=1}^{N}S(i,delay(x,y,i)) \tag{2}$$

where $I_{DAS}(x,y)$ represents the pixel value at position $(x,y)$ within the imaging region. $N$ denotes the number of sensors. $S(i,delay(x,y,i))$ represents the data value in the sensor data received by the $i^{th}$ transducer element with a delay of $delay(x,y,i)$. After traversing the above calculations over the entire imaging plane, we can obtain the reconstructed image.

### 2) Forward Model: s-Wave Algorithm

s-Wave is a forward simulation method for obtaining PA signals from images. In comparison to the MATLAB k-Wave toolbox [29] which is able to simulate PA imaging numerically, s-Wave significantly improves simulation speed while ensuring that the results closely resemble the signals from k-Wave [23]. We found that each pixel value in a PA image generates a set of signals after the forward process, and the shapes of the signals obtained at different pixels are very similar, mainly differing only in magnitude and phase. The differences can be characterized by the spatial relationship between pixels and sensors.

Initially, s-Wave utilizes k-Wave to obtain a set of standard sensor data, denoted as $s\in\mathbb{R}^{1\times M}$ (M is the sampling depth), which is acquired from unit pixels at the center of the imaging region. Then, the phase will change across the entire time domain range of the signal based on the offset as expressed as

$$\tau(d_p,d_s)=\frac{d_p-d_s}{c}\cdot f_s \tag{3}$$

where $\tau(d_p,d_s)$ represents the delay difference between the other pixel and the central pixel position. The $d_p$ denotes the distance from the specified pixel to each sensor, and $d_s$ is the distance from the central pixel to each sensor. The amplitude coefficient is related to the intensity of pixel values and the energy attenuation in the medium so that the coefficient can be defined as

$$A(d_p;p)=k\frac{p}{{d_p}^2} \tag{4}$$

where $A(d_p;p)$ is the amplitude coefficient for the specified pixel position. Here, $k$ is a constant, and $p$ represents the pixel value. Therefore, the signal $s_p\in\mathbb{R}^{1\times M}$ obtained at the specified pixel point can be expressed as

$$s_p=A(d_p;p)\cdot\langle s|\tau(d_p,d_s)\rangle \tag{5}$$

where $\langle\cdot|\cdot\rangle$ denotes a loop operator, which means changing the phase of each value in $s$. By traversing all pixels, we can get many groups of signals based on $s$. The number of groups is the same as the number of pixels. By superimposing these signals together, we can get a new set of signals $s_{np}\in\mathbb{R}^{1\times M}$. Then we traverse all the sensors and repeat the above operation to obtain the new PA signal $s_n\in\mathbb{R}^{N\times M}$ ($N$ is the number of sensors). The specific description is shown in Algorithm 2.

**Algorithm 2** s-Wave

**Require:** Input PA image $I$; Standard sensor data $S$; the reference distance $d_s$; the number of sensors N; ROI size $n\times m$; pixel position $(x_p,y_p)$; sensors position $(x_s,y_s)$; Sound speed c; the sample frequency $f_s$; constant $k$.
**Ensure:** New sensor data $s_n$.

```
for i ← 1, N do
    for j ← 1, n × m do
        d_ij = sqrt((x_pj − x_si)^2 + (y_pj − y_si)^2)
        τ(d_ij, d_s) = (d_ij − d_s)/c · f_s
        A(d_ij, I_j) = k I_j / d_ij^2
        s_ij = A(d_ij; I_j) · ⟨s|τ(d_ij, d_s)⟩
    end for
    s_n^(i) = Σ_{j=1}^{n×m} s_ij
end for
s_n = Σ_{i=1}^{N} s_n^(i)
return Reconstructed image s_n
```

### 3) Model-based PA Image Reconstruction Algorithm

The model-based algorithm is an iterative image reconstruction method, in which DAS serves as the backward model from signals to images, and s-Wave acts as the forward model from images to signals [18], [19]. The specific description is shown in Algorithm 3.

**Algorithm 3** Model-based Reconstruction with s-Wave.

**Require:** Input sensor data $S$; max iteration number $K$; initial learning rate $lr$; loss threshold $L$.
**Ensure:** Final reconstructed image $I_t$.

```
I_0 = DAS(S)                         ▷ Get the initial reconstructed image
s_n = sWave(abs(I_0))                ▷ Get the new sensor data
R_1 = S − s_n                        ▷ Compute the difference
for t ← 1, K do                      ▷ Iterate and get the final image
    I'_t = DAS(R_t)
    I_t = I_{t−1} + lr * I'_t        ▷ Weighted superposition
    R_{t+1} = S − sWave(abs(I_t))    ▷ residual sensor data
    if ||R_{t+1}||_2 < L then        ▷ threshold judgment
        return
    end if
end for
return Reconstructed image I_t
```

Initially, the algorithm processes the raw sensor data $S$ using the DAS method to produce an initial reconstructed photoacoustic image $I_0$. This image $I_0$ is then converted to its absolute value and fed into the s-Wave, which generates new sensor data $s_n$. The difference between the original sensor data $S$ and the new sensor data $s_n$ results in the residual sensor data $R_1$. Subsequently, the DAS uses $R_1$ to create a residual image, which is scaled by a learning rate $lr$ and added to the previously reconstructed image from the DAS. The combined image is then reintroduced into the s-Wave, where its output is once again differenced with the original sensor data $S$ to yield the updated residual data $R_{t+1}$. During each iteration, a loss value is computed based on $R_{t+1}$. If this loss falls below a predetermined threshold, the iterative process terminates, and the most recent combined image will be output as the final result. Otherwise, iterations continue until the maximum number of iterations, predefined at the start, is reached.

This iterative approach ensures that each subsequent image reconstruction refines the previous one, enhancing the accuracy and quality of the final photoacoustic image. By dynamically adjusting and converging towards the optimal image reconstruction, this method leverages both the DAS and s-Wave algorithms in a complementary manner, providing a robust solution for real-time, FPGA-accelerated photoacoustic tomography.

## C. Hardware Architecture

### 1) Architecture Overview

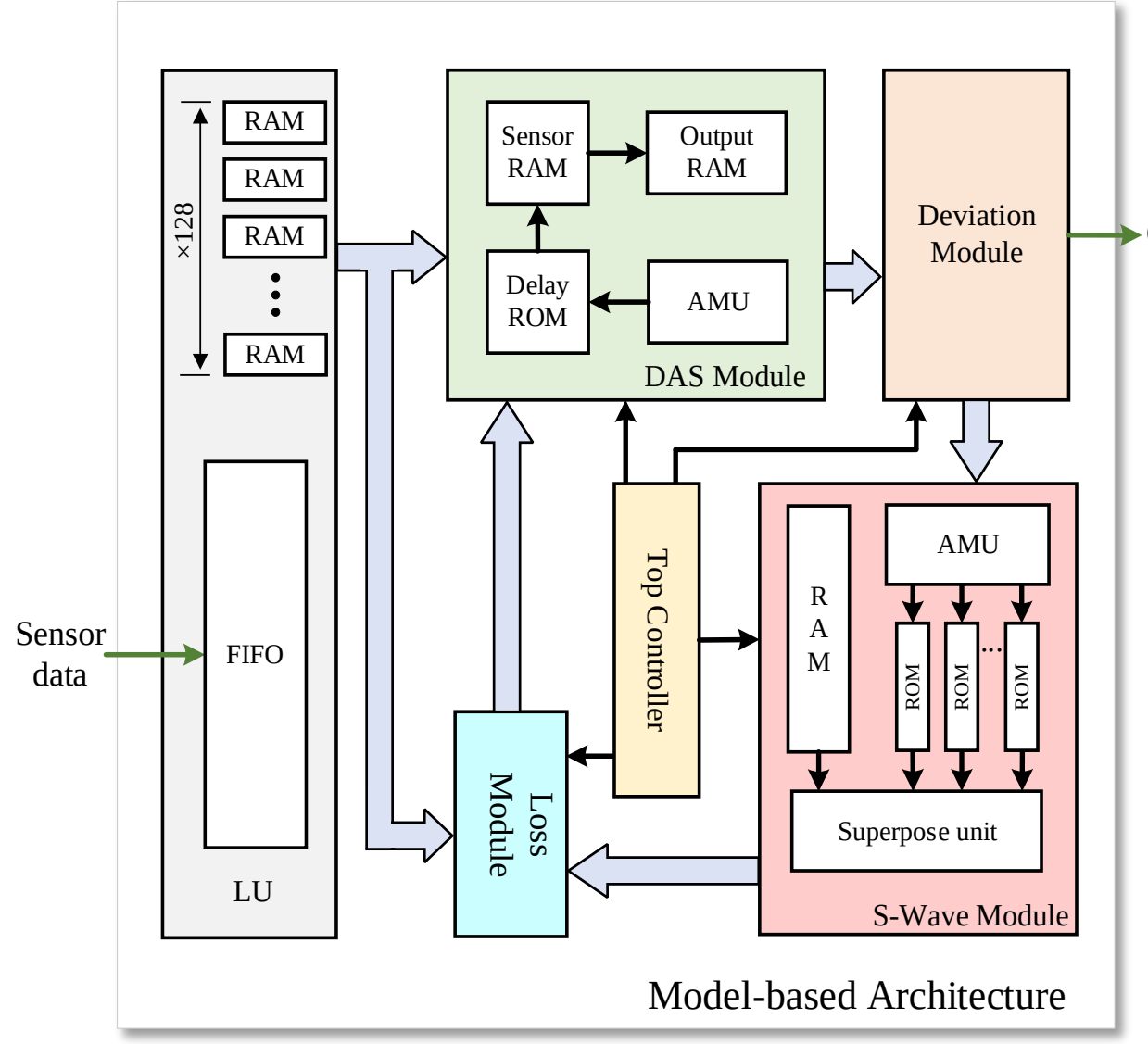

Fig. 2. Hardware architecture of model-based algorithm with s-Wave. FIFO: first in first out; LU: load unit; AMU: address mapping unit.

The hardware acceleration architecture of model-based algorithm consists of 6 sub-modules, namely, the LU, DAS module, Deviation module, s-Wave module, Loss module, and Top Controller module, each represented in different colors as shown in Fig. 2. In the system, the sensor data is ingested as the input and processed to produce the reconstructed image as the output. The LU is tasked with buffering the incoming raw sensor data and ensuring its orderly storage. The DAS module then processes this data by employing the delay-and-sum algorithm to effectively extract and aggregate the sensor data into a corresponding PA image, which is subsequently outputted. The Deviation module focuses on calculating the difference between successive iterations to generate a residual image and is also responsible for outputting the final image upon completion of the model-based algorithm iteration number. Meanwhile, the s-Wave module performs the forward transformation of the image into sensor data and passes it to the Loss module, where the sensor data residuals and the final loss value are computed. The Top Controller module controls the behavior of each sub-module according to the execution status of the system.

### 2) Data Reuse Method

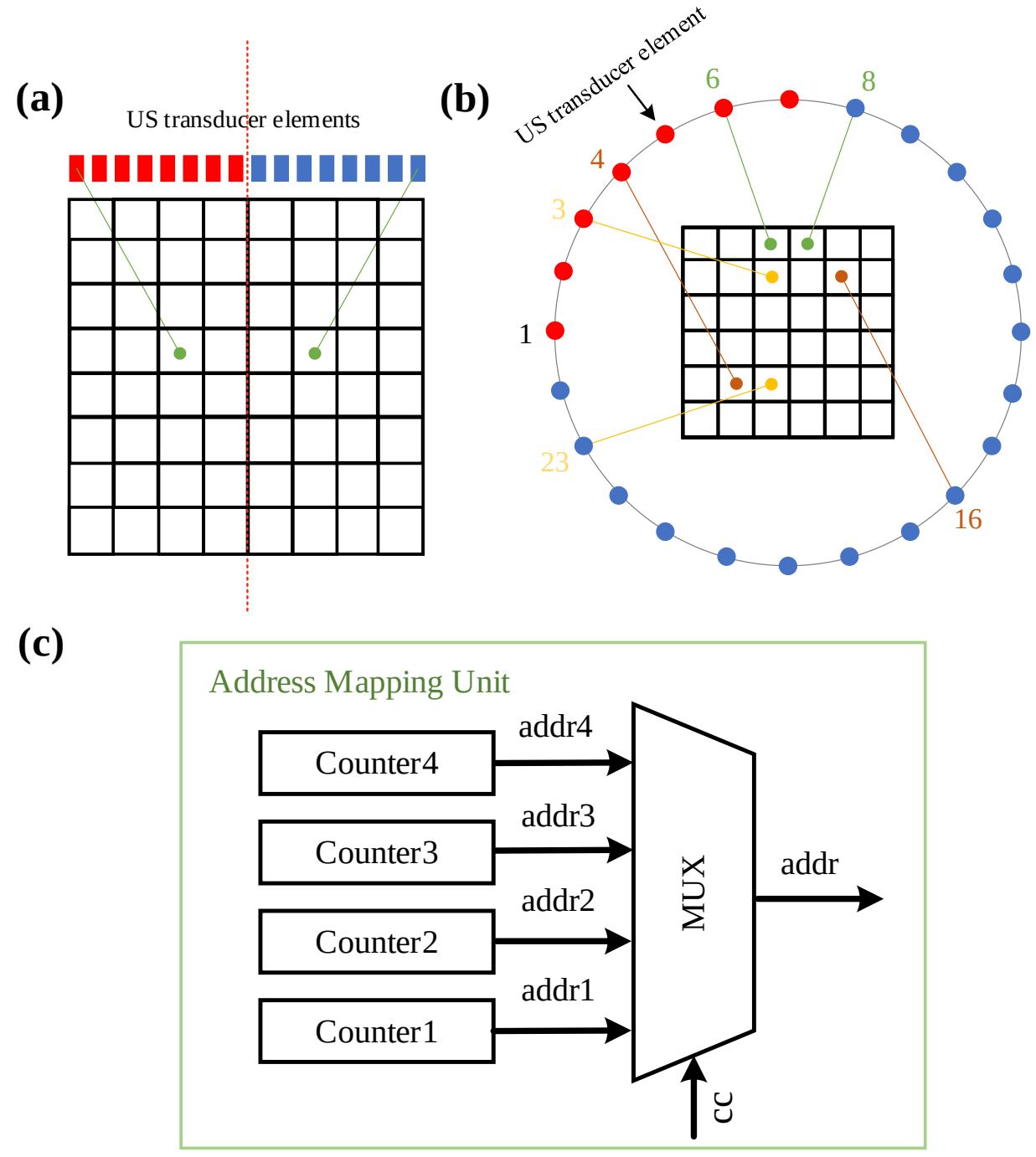

Fig. 3. Proposed data reuse method. (a) Symmetry description of linear transducer array and imaging region. (b) Symmetry description of ring-shaped transducer array and imaging region (c) Hardware architecture of address mapping unit (AMU). cc: execution cycle count.

In a PA imaging system, the transducer elements commonly arranged around the imaging target often exhibit a certain degree of symmetry. Examples include the linear arrays, semi-circular arrays, circular arrays, and even spherical arrays that we frequently use. This indicates that using popular symmetric array configurations can often yield better results, both at the signal and image levels. When implementing model-based algorithms in hardware, it is necessary to pre-store the delay data from the DAS algorithm, as well as the phase shift data and amplitude weighting data from the s-Wave algorithm. For instance, assuming 128 transducer elements and an ROI size of $128 \times 128$, the storage size required for the delay data (with a precision of 10 bits) is $128 \times 128 \times 128 \times 10$ bits $\approx$ 21 Mb. Similarly, the storage size for phase shift data (with a precision of 10 bits) is $128 \times 128 \times 128 \times 10$ bits $\approx$ 21 Mb and the storage size for amplitude weighting data (with a precision of 8 bits) is $128 \times 128 \times 128 \times 8$ bits $\approx$ 17 Mb. Altogether, storing these parameters demands approximately 60 Mb of on-chip storage resources, which imposes a significant burden on the storage capacity of FPGA. For 3D PA imaging, this will become much more challenging, requiring several Gb storage resources.

Our proposed data reuse method significantly alleviates the storage burden. Consider the linear array arrangement in Fig. 3(a), where both the array elements and the ROI are symmetrical about the central axis. The the first element on the left and the last element on the right are symmetrical. In addition, the line segments of the same color have equal lengths which means the equal distances. Therefore, the distances from the last element to all pixel positions in the ROI can be derived from the corresponding distances from the first element to all pixel positions due to this symmetry. In this way, we only need to store the parameters corresponding to the red array elements on the left side. Correspondingly, the circular array shown in Fig. 3(b) can be analyzed to exploit symmetry for data reuse. Taking a 6×6 imaging area surrounded by 24 sensors arranged in a circular array as an example, the two sensors indicated by the line segments of the same color are symmetric with respect to the origin point. This symmetry allows us to selectively store data for a subset of sensors, which is red as illustrated in Fig. 3(b), efficiently reducing storage requirements. Extrapolating this concept to 128 sensors which is also what we actually used in this paper, we only need to pre-store data corresponding to 17 sensors at a minimum, which is more than 7 times storage reduction. Considering the trade-off between area and speed, we decided to pre-store the parameters for 33 sets of sensors, which resulted in more than 3 times resource savings. This optimization allowed us to implement a 32-channel data parallel strategy within the FPGA, significantly enhancing the efficiency and performance. The implementation of the address multiplexing in hardware is accomplished by configuring 4 constraint-based counters which also implies four addressing modes and a multiplexer in Fig. 3(c) following center-symmetry rules to construct an address mapping unit (AMU) for data access during various execution cycle (for a total of 128 channels of data, with 32 channels processed in parallel, $cc \leqslant$ 128/32=4).

### 3) Hardware Implementation of DAS

In the DAS implementation circuit as shown in Fig. 4, we pre-store delays in 33 read-only memories (ROMs) for each sensor channel, with a delay data size of 128×128 to match the image size (the image grid size of the ROI region is set to 128×128). When the sensor data in the LU becomes available, it is fed to the DAS module as soon as possible. To enhance data processing efficiency, we employ a 32-channel parallel data processing strategy, requiring 4 execution cycles to process the 128 channels of sensor data. Utilizing the data reuse method previously discussed, we enable 32 out of the 33 delay ROMs for each execution cycle. Specifically, when $cc$ =1 or 3, ROM0-ROM31 are enabled, and when $cc$ =2 or 4, ROM1-ROM32 are enabled. Additionally, the AMU unit adjusts the addressing mode of the delay ROM based on the execution cycle to ensure that the correct delay information is matched with the corresponding sensor channels.

Once the sensor data streams into the RAM, the DAS module begins addressing the ROM to retrieve the delay values. These delay values serve as addresses to fetch the corresponding delayed sensor values from the RAM. We employ a three-stage pipeline to sum the retrieved sensor values. The summed result is added to the corresponding position in the RAM, which stores the final image (initial value = 0). During the last execution cycle, the written image values are taken as absolute values and written back to the same positions. This accumulation process involves simultaneous read and write operations in the RAM, with the write data stream trailing the read data stream by two clock cycles.

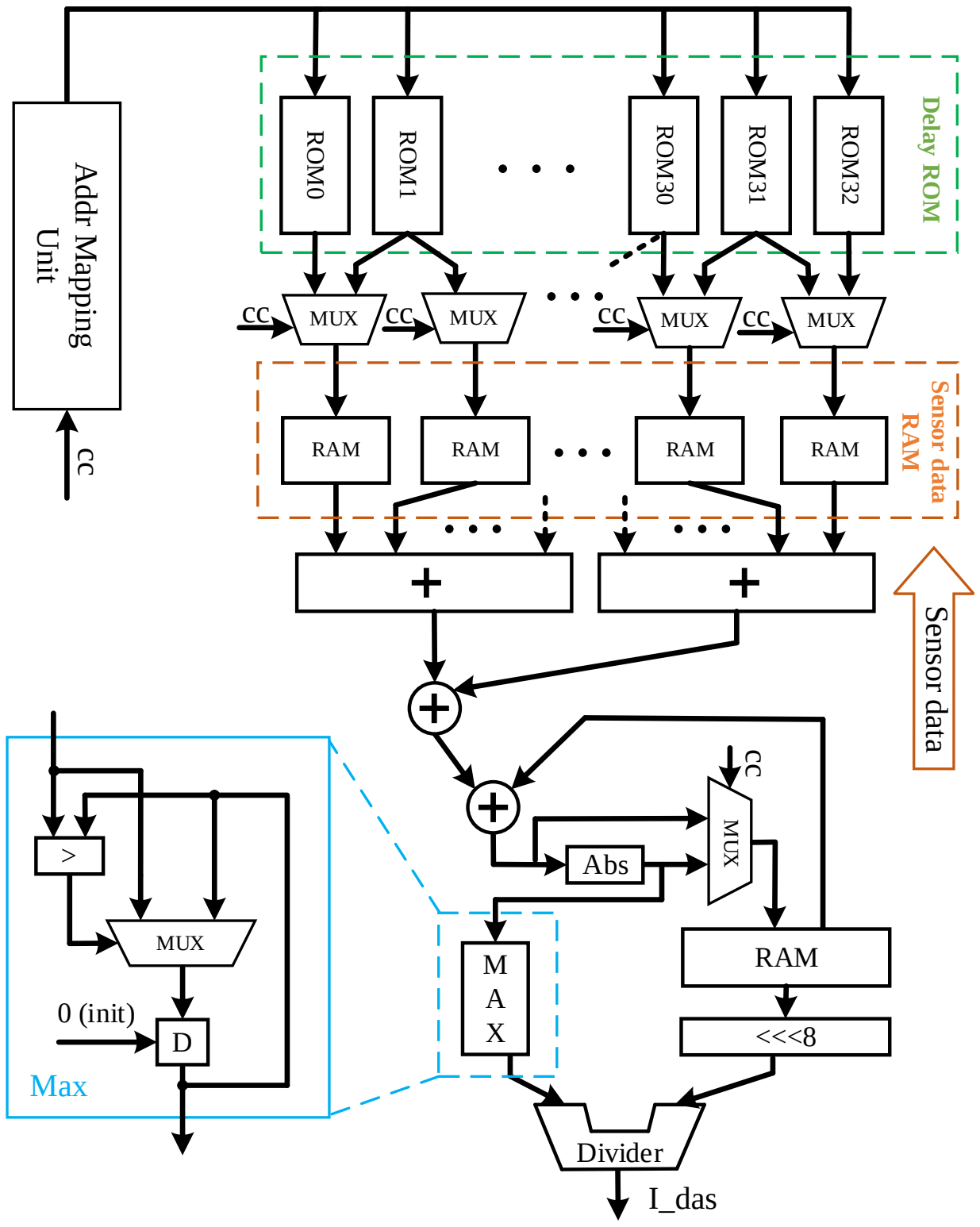

Fig. 4. Structure of the DAS module. D: register.

To prevent unnecessary large data values caused by subsequent iterative algorithms, we can normalize the entire image, constraining the image values within the range of 0 to 256. The Max unit is responsible for finding the maximum value in the whole image while writing the image values during the final execution cycle. It consists mainly of a comparator, a selector, and a local register array (initial value = 0). After the DAS image accumulation is completed, the pipeline output is left-shifted by 8 bits to achieve a 256-fold magnification then fed into a divider along with the maximum value to perform normalization.

### 4) Hardware Implementation of Deviation Module

As illustrated in Fig. 5, the input of the Deviation module is an image generated from the DAS. Inside the module, the data flows along two paths. One path applies a learning rate $lr$ weighting to the image, subtracts it from the previous iteration image stored in RAM, computes the absolute value, and then inputs it into a multiplexer. The other path directly inputs the image into another port of the multiplexer. The multiplexer selects its output based on the iteration count $t$ of the model-based algorithm. When $t = 0$, the DAS image is directly output from the multiplexer and written into the RAM. For $t > 0$, the multiplexer outputs the residual between the two iterative images and writes it into the RAM, overwriting the previous image.

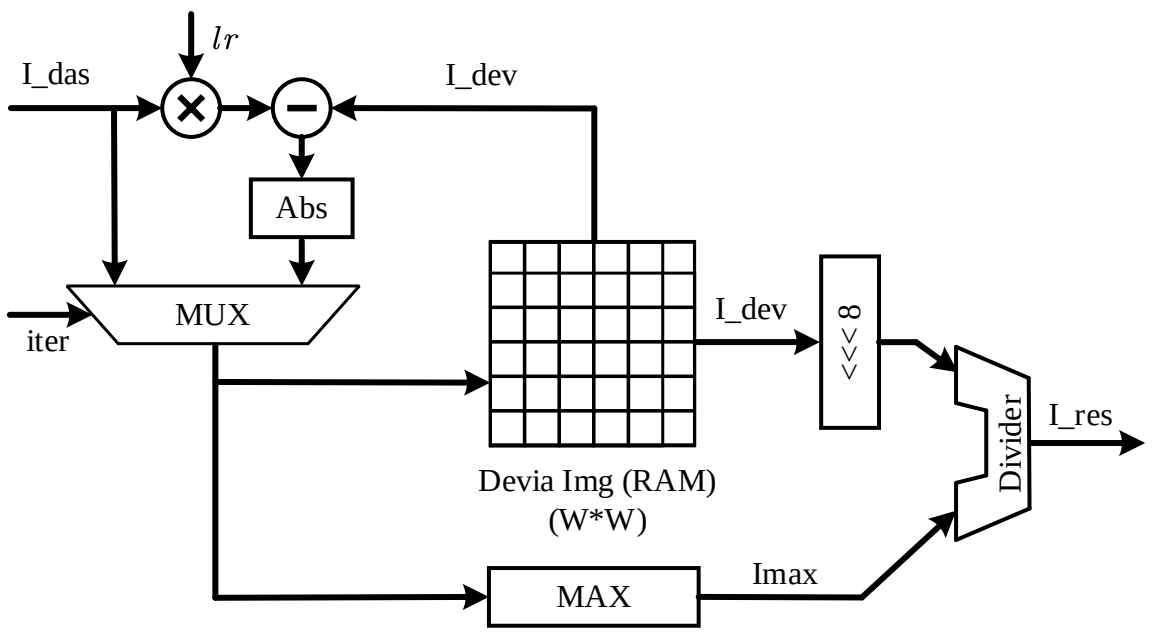

Fig. 5. Structure of the Deviation module.

Once the residual image is written, it is output from the module as quickly as possible. During this process, the image range is constrained to 8 bits using circuitry similar to the output design of the DAS module. The direction of the output depends on the status of the iterative algorithm. If the iteration has not ceased, the output is fed into the s-Wave module for further processing. Otherwise, the Deviation module outputs the final reconstructed image. All the aforementioned processing is performed using a pipelined approach to increase the throughput of the module.

#### 5) Hardware Implementation of s-Wave

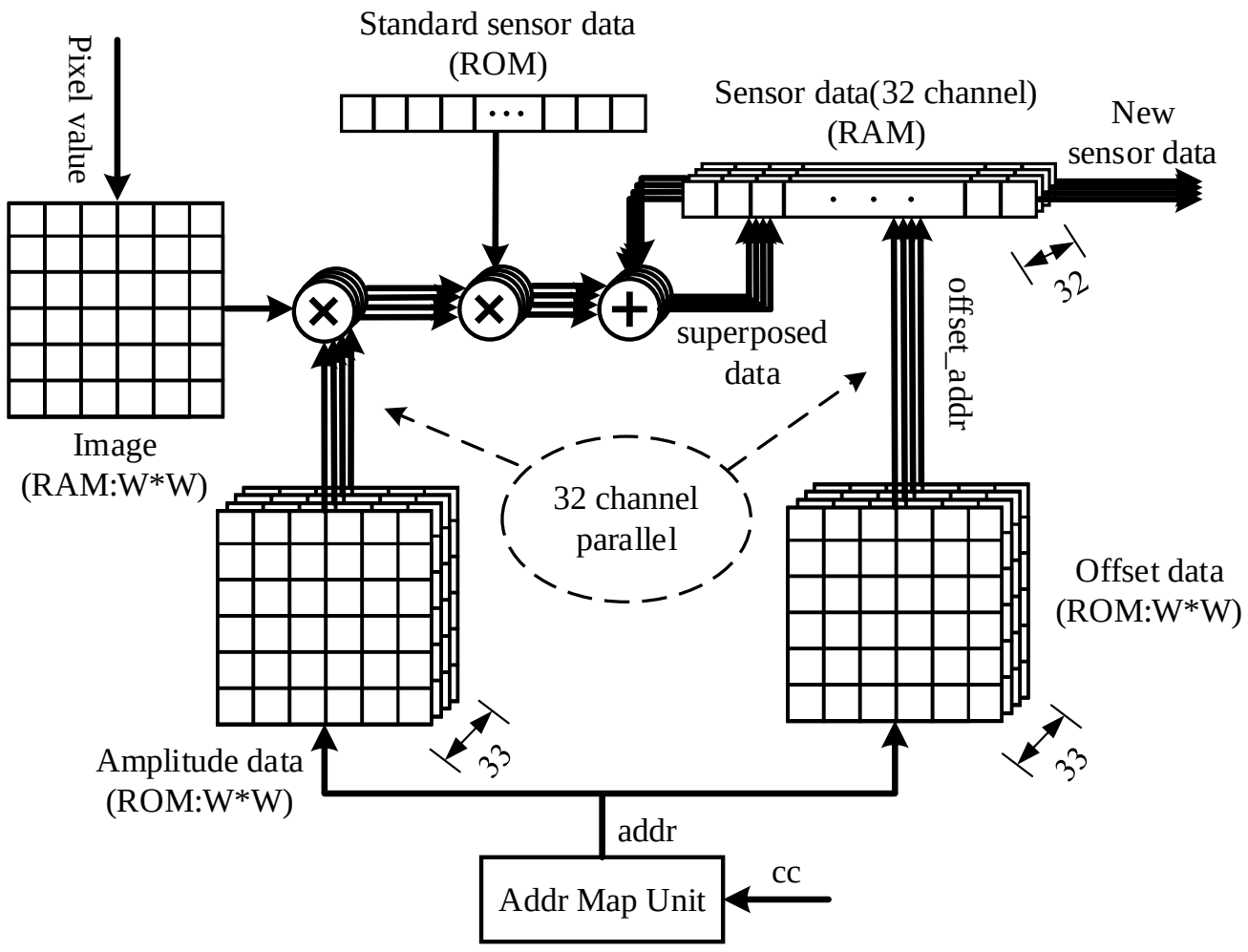

Fig. 6. Structure of the s-Wave module.

The s-Wave module, as illustrated in Fig. 6, is designed with preloaded 33 sets of amplitude weighting data, 33 sets of offset data, and a single set of standard sensor data $s$ stored in ROMs. Upon the complete transfer of pixel values from the Deviation module into the RAM, the s-Wave module initiates its processing sequence. Initially, pixel values are sequentially retrieved, simultaneously with the 32 amplitude values fetched from the ROMs. The enablement of 32 out of the 33 ROMs is determined by the current execution cycle. The pixel value is then broadcasted and multiplied by the 32 amplitude values, an operation that is efficiently completed within one clock cycle. These resultant values are then weighted and subsequently added to the standard sensor data, which has been broadcasted from the ROM, effectively modifying its amplitude characteristics. Next, the phase alteration is achieved by adjusting the start and end positions of data writing in the RAM. Specifically, the 32 phase offset data are read from the ROMs and utilized as read addresses to fetch the corresponding sensor data from the 32 RAMs. The amplitude-weighted standard sensor data are then summed with these retrieved sensor data corresponding to the offset data, and the results are stored back in their original RAM positions. During this accumulation phase, the generation of RAM write addresses is achieved by simply delaying the offset data by three clock cycles using a local register array. The ROM-stored amplitude and offset data are addressed correctly through the AMU, capable of generating four distinct addressing modes, analogous to those used in the DAS module. After all pixel values within the input image are processed, the s-Wave module completes the generation of 32 new sensor data sets, achieving full output of all 128 channels of new sensor data within four execution cycles.

It is crucial to highlight that, for each individual pixel value, we must perform amplitude weighting across the entire length of the standard sensor data, followed by a systematic accumulation of data from RAM. This intricate processing sequence involves an unrolling procedure, necessitating the use of a finite state machine (FSM) to manage the conditional state transitions effectively. By employing an FSM, we can meticulously orchestrate the various stages of the data processing pipeline, namely, the fetching, weighting, and accumulation of sensor data, ensuring that each operation adheres to both logical coherence and timing precision. Each state in the FSM corresponds to a specific phase of this pipeline, with state transitions dictated by predefined conditions based on execution cycles and data status. This method enhances the logical structure and timing efficiency of the data processing. The FSM framework provides granular control over the data processing flow, with each state representing a distinct phase of the pipeline.

#### 6) Hardware Implementation of Loss Module

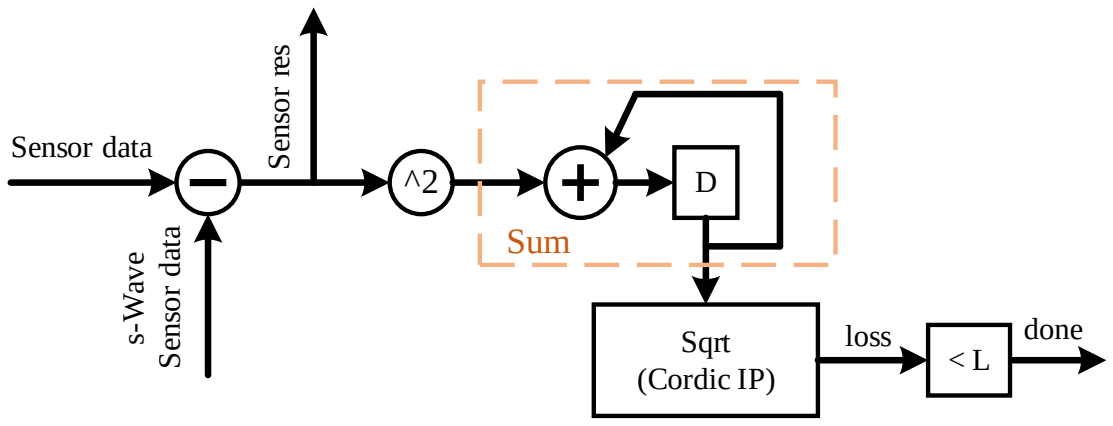

Fig. 7. Structure of the Loss module.

The Loss module, shown in Fig. 7, is tasked with computing the difference between the newly generated sensor data from the s-Wave module and the originally measured sensor data to derive the sensor residuals, which are then squared and accumulated using an adder and a register to compute the sum of squares. Subsequently, the accumulated result is subjected to a square root operation using a CORDIC (Coordinate Rotation Digital Computer) algorithm IP core to obtain the final loss value. This loss value is then compared with a predefined threshold to determine whether the algorithm iteration should terminate, generating an iteration end signal based on this comparison.

## III. Experimental Results

To demonstrate the feasibility of our proposed hardware architecture for accelerating the model-based algorithm with s-Wave and its potential integration into a palm-sized PAT system, we conducted experiments using both phantom and in vivo human finger data on an FPGA platform.

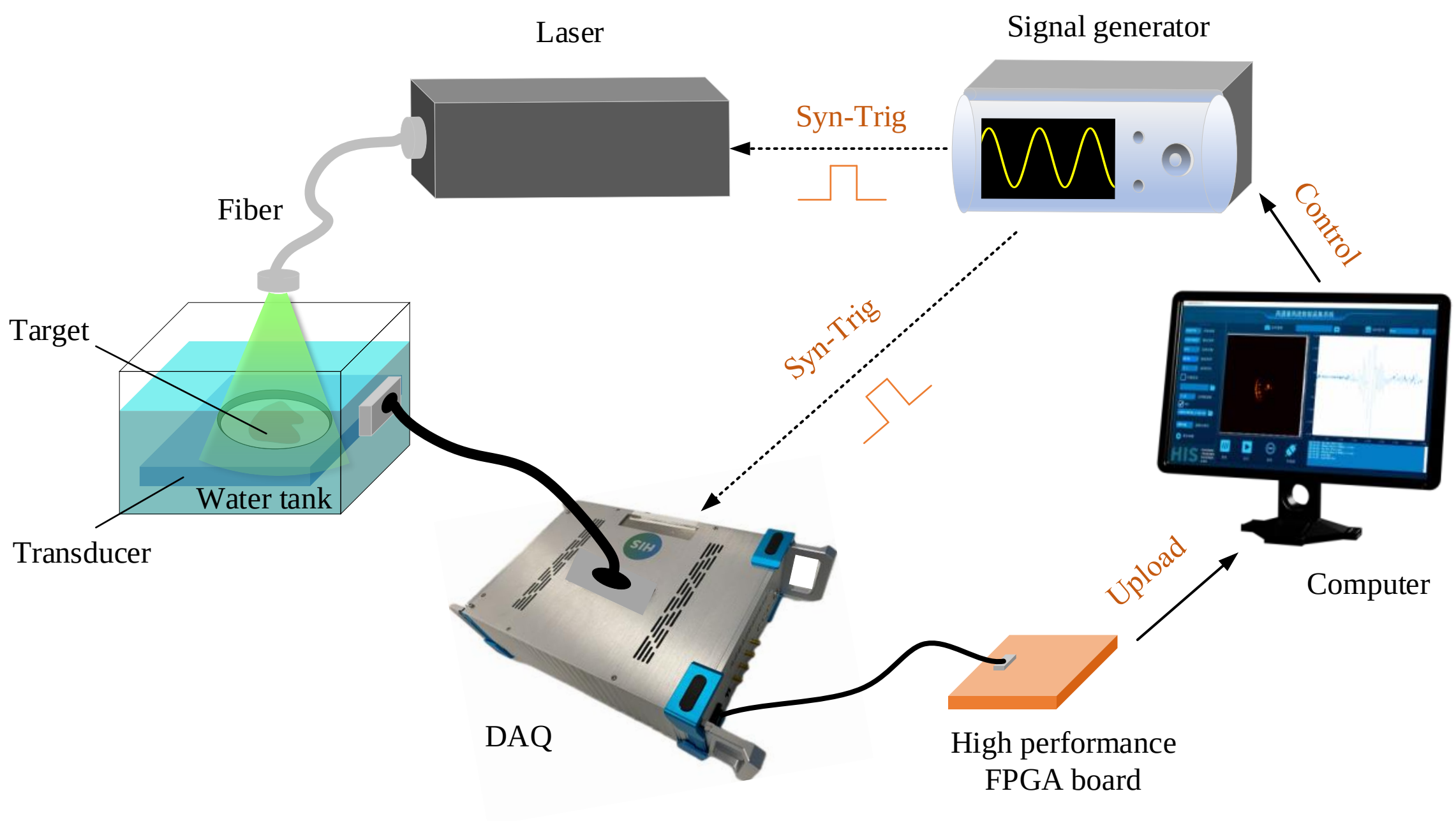

Fig. 8. The experiment setup of PA imaging system. DAQ: data acquisition card.

### A. Experimental Setup

In our experiments, the PA signals were captured using a 128-element ring-shaped ultrasound transducer array with a radius of 30mm [30], [31]. Data acquisition was handled by an in-house developed data acquisition card (DAQ) capable of supporting 256-channel signal acquisition with a maximum sampling frequency of 80 MSPS and a maximum sampling depth of 4096 samples. The FPGA platform utilized is the Xilinx AXU15EG MPSOC, featuring a 200 MHz clock frequency, 341280 Look-up tables (LUTs), and 26.2 Mb block RAM. As illustrated in Fig. 8, the experimental system is orchestrated such that the PC controls the signal generator to emit synchronized pulse signals to both the laser and the DAQ. The laser beam, directed through an optical fiber, targets the object located at the center of the ring transducer, generating PA signals that are subsequently received by the transducer. These signals are then transmitted to the DAQ, where they undergo amplification, filtering, and other preprocessing steps before being fed into the FPGA for image reconstruction. The reconstructed images are finally uploaded to the PC for display. The selected Region of Interest (ROI) (128×128 grids) measures 20 mm × 20 mm.

### B. Implementation Results

#### 1) Reconstructed image results

We obtained three sets of PA data by experiments: 1) cross-sectional data three pencil leads placed perpendicular to the plane of the ring-shaped transducer; 2) cross-section data of two pencil leads placed parallel to the plane of the transducer; 3) and *in vivo* human finger data. We employed hardware implementations of the DAS algorithm, the DMAS algorithm, the DAS-CF algorithm, and our proposed model-based algorithm to obtain image reconstruction results for three sets of data, as illustrated in Fig. 9. Furthermore, we compared the results between the hardware implementation of the model-based algorithm and the original software-version algorithm. The reconstructed images reveal that the DAS algorithm yields relatively coarse results, marred by significant noise and artifacts. However, both the DMAS and DAS-CF algorithms demonstrate substantial improvements over the DAS algorithm across all three datasets, as evidenced by their superior suppression of background noise, enhancement of the primary PA signal, and effective reduction of sidelobes. Compared to these DAS-based methods, the proposed novel hardware architecture of iterative model-based algorithm not only effectively eliminates background noise and artifacts, but also significantly suppresses interference around and within the imaging targets. Additionally, it enhances the relative intensity of key pixels in the imaging targets, thereby greatly improving image contrast.

Detailed comparisons in the blue box, as illustrated in Fig. 9, show that in the reconstructed images of the three pencil leads, the contrast intensity of the two pencil lead sections within the white dashed box is noticeably lower in the DMAS and DAS-CF results compared to the model-based algorithm. This indicates a loss of some effective signals in the former two methods while suppressing sidelobes. For the results involving two pencil leads, although the DMAS and DAS-CF images exhibit a significant reduction in noise, a substantial amount of artifacts persists around the imaging targets, as shown by the white dashed box. In stark contrast, the results of our proposed method are virtually devoid of noise and artifacts. This trend is similarly observed in the finger images. It is particularly noteworthy that the DMAS-generated finger images display numerous additional, albeit slightly weaker, pixels around some blood vessels, which can be attributed to spatial correlations among the image pixels. Conversely, the results from the DAS-CF and model-based algorithms are comparatively superior, as they exhibit fewer interfering pixels around the blood vessels, thereby facilitating a clearer distinction between different vessels. Additionally, the model-based algorithm results, in contrast to those from the DMAS and DAS algorithms, enhance the contrast of specific pixels corresponding to the capillaries on the skin surface, thereby rendering the boundary of the finger more distinct.

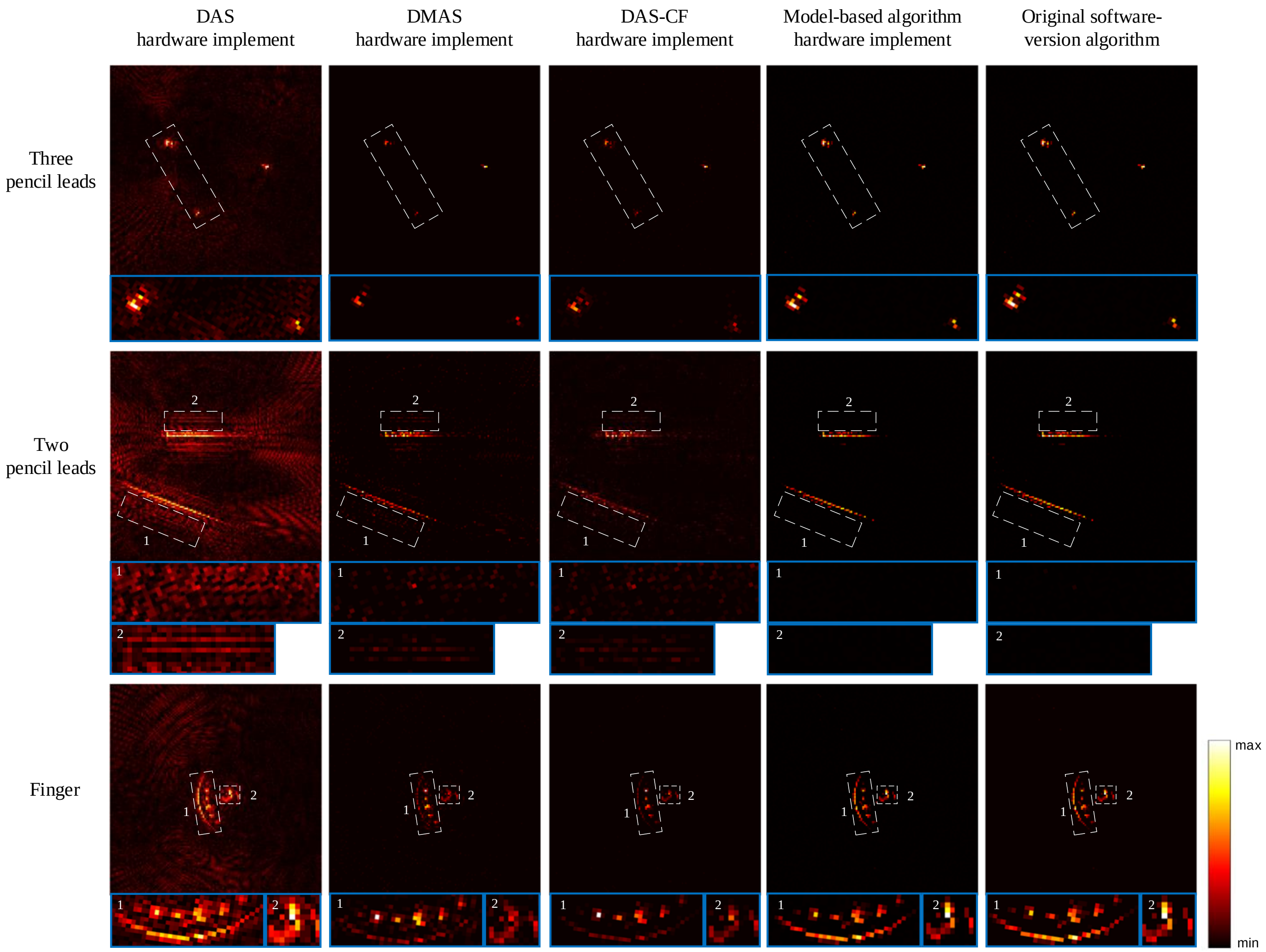

Fig. 9. Comparison of image reconstructed results from DAS (hardware implementation), DMAS (hardware implementation), DAS-CF (hardware implementation), the Model-Based Algorithm (hardware implementation) and the original Model-Based Algorithm (software implementation). The datasets used from top to bottom are: three pencil leads perpendicular to the transducer plane, two pencil leads parallel to the transducer plane, and an in vivo human finger. White dotted box: the chosen regions of interest (ROI). Blue box: zoomed ROI.

In addition, we conducted a comparative analysis between the hardware implementation results of the model-based algorithm and those derived from the original software algorithm, as illustrated in Fig. 9, where it is apparent that the results from both implementations are nearly indistinguishable. To further investigate the differences between these two sets of results, we present the error maps between the hardware implementation and the original software algorithm, as shown in Fig. 10. We can see that the error map of the hardware implementation of the DAS algorithm exhibits significant discrepancies compared to the results of the original software version model-based algorithm. This substantial difference can primarily be attributed to the pronounced background noise and artifacts present in the DAS results. Conversely, the hardware implementations of the DMAS and DAS-CF algorithms reveal discrepancies with the software-based model results mainly in terms of pixel intensity within the target object, as well as artifacts surrounding and within critical pixels adjacent to the target object. These error maps of the hardware and software results of model-based algorithm reveal that while there are a few individual pixels diverge, these discrepancies are primarily due to the data quantization from floating-point to fixed-point within the FPGA. Notably, even the lower intensity pixel clusters in the error maps continue to delineate the contours of the target objects, indicating that the differences between the hardware and software implementations are largely confined to variations in pixel intensity. This observation suggests that the performance optimization achieved by the hardware implementation closely mirrors that of the original software algorithm, which affirms the fidelity of the hardware implementation. Furthermore, it implies that the differences in pixel intensity can be effectively addressed through appropriate scaling adjustments, thereby ensuring the robustness and reliability of the hardware implementation.

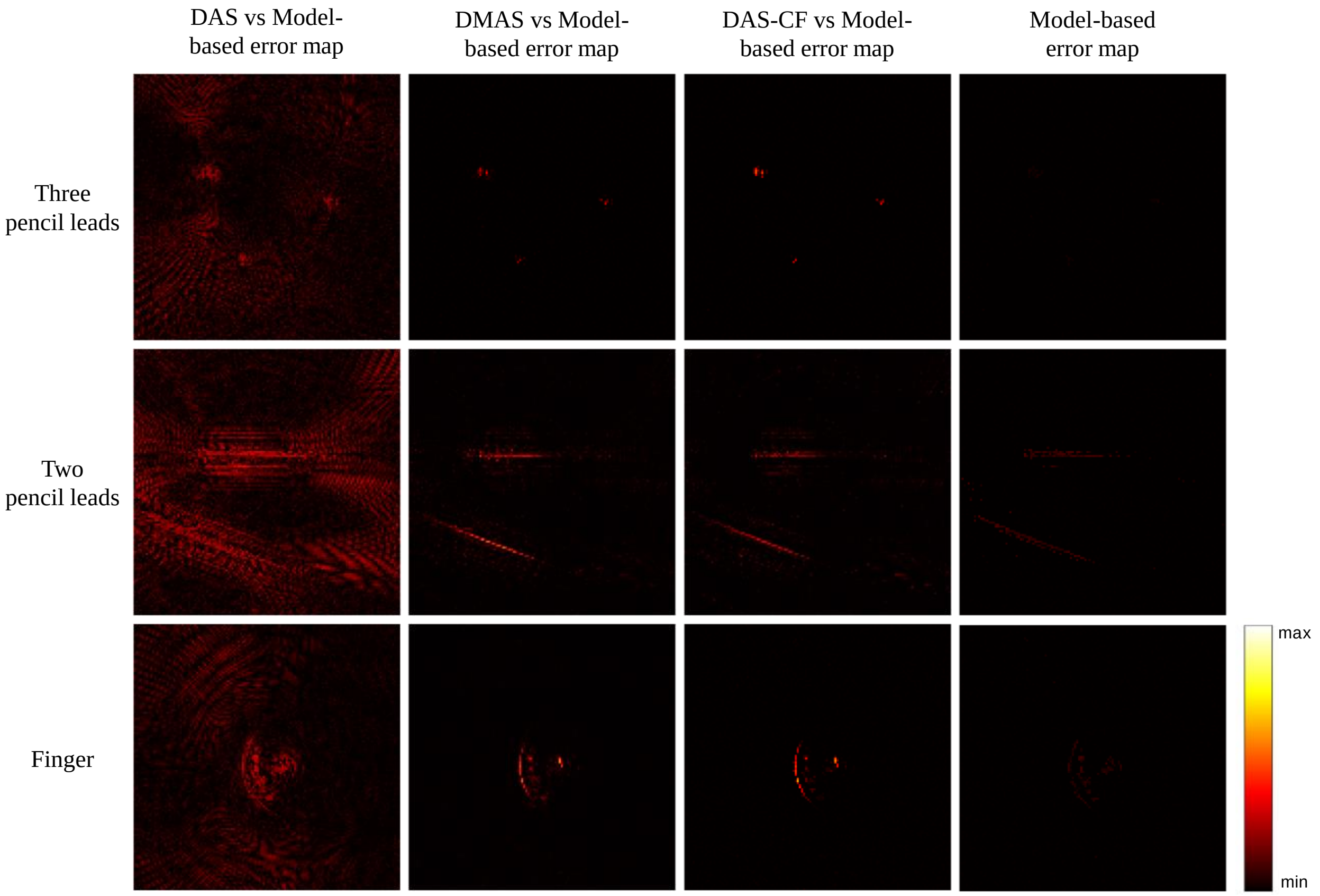

Fig. 10. The error maps of hardware implementations of DAS, DMAS, DAS-CF, and Model-based algorithms compared to the original software-version model-based algorithm results.

We computed the structural similarity (SSIM) between the hardware-reconstructed images (DAS, DMAS, DAS-CF, Model-based) and those reconstructed by the original software-version model-based algorithm. As shown in Fig. 11, the SSIM values reveal that the DAS algorithm exhibits relatively poor performance, while the DMAS algorithm shows only average results. The DAS-CF algorithm performs reasonably well; however, its effectiveness markedly declines in the presence of substantial significant noise and artifacts. The SSIM values for the images of two pencil leads, three pencil leads, and an in vivo human finger reconstructed by model-based algorithm were 0.9466, 0.9734, and 0.967, respectively, with an average SSIM reaching 0.9623. This high level of structural similarity further corroborates the observation that the results obtained from the hardware implementation are exceedingly close to those produced by the original software algorithm.

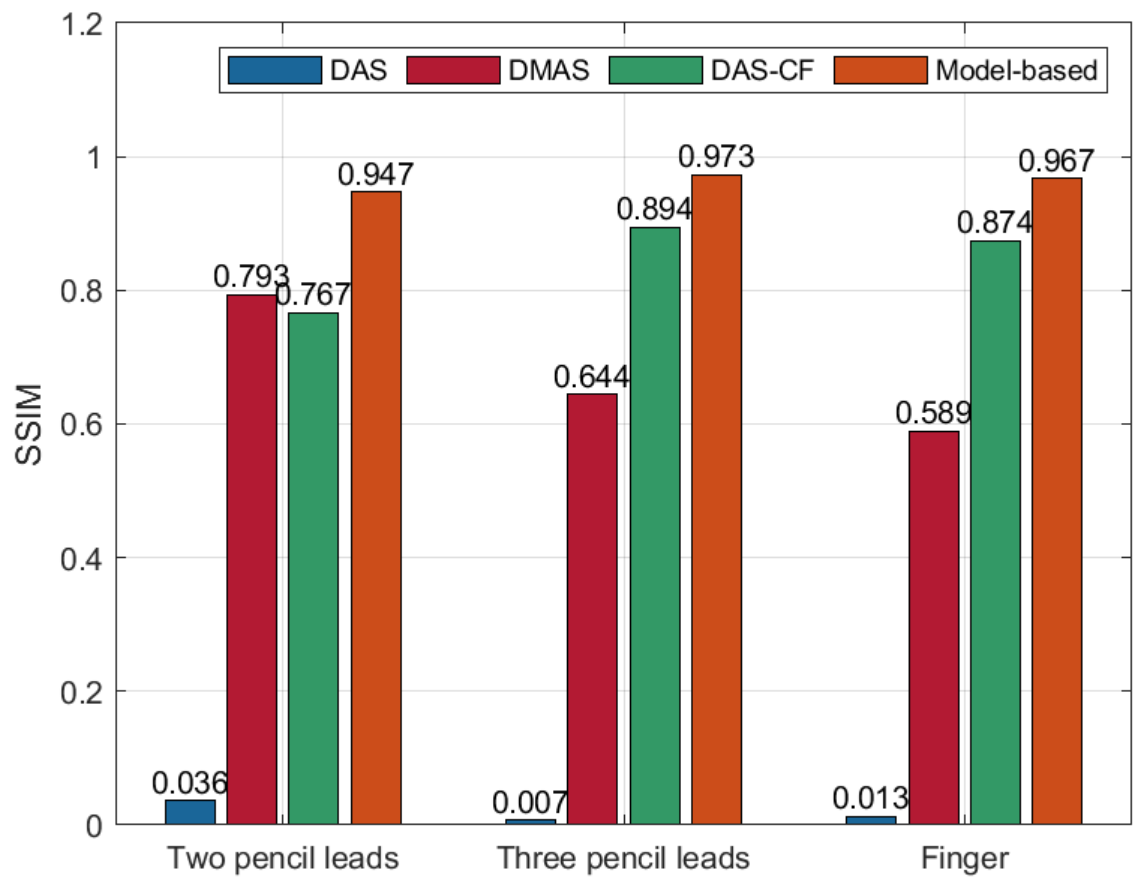

Fig. 11 SSIM of the results generated by the hardware implementation of DAS, DMAS, DAS-CF and the Model-based algorithm, comparing to the original software algorithm results.

### 2) Hardware implementation performance

Given that the original model-based software algorithm utilizes floating-point arithmetic, which is prohibitively expensive for hardware implementation so that all algorithmic operations are employed by fixed-point quantization on the FPGA. Since this paper represents the first implementation of a model-based algorithm on hardware, we compared the performance of our proposed design with the original software algorithm running on a CPU, as detailed in Table I. The original algorithm was developed using MATLAB R2021 and executed on a personal computer (PC) equipped with an Intel i5 12500H CPU and 16 GB memory. The performance comparison results showcased in TABLE I reveal that while the CPU takes several seconds to tens of seconds to reconstruct one frame, the FPGA implementation achieves speeds approximately 26 to 270 times faster. This speed advantage is even more pronounced for more complex imaging targets. The resource utilization of the FPGA

TABLE I
PERFORMANCE COMPARISON

| Platform | Intel i5 12500H | | | Xilinx AXU15EG | | |
|---|---|---|---|---|---|---|
| Clock (MHz) | N/A | | | 200 | | |
| Time (one frame) | Three encil leads | Two pencil leads | Finger | Three encil leads | Two pencil leads | Finger |
| | 2.054s | 17.576s | 12.715s | 0.0795s | 0.1332s | 0.0465s |
| Speed (FPS) | 0.49 | 0.06 | 0.08 | 12.58 | 7.5 | 21.5 |

is detailed in TABLE II, with the corresponding allocation of major modules illustrated in Fig. 12.

Additionally, we compared the power consumption of the CPU and FPGA, as shown in Fig. 13. The power consumption of the FPGA is more than 10 times lower than that of the CPU, resulting in an energy efficiency improvement of up to approximately 2700 times.

TABLE II
RESOURCE UTILIZATION

| Resource | LUT | LUTRAM | FF | BRAM | DSP |
|---|---|---|---|---|---|
| Used | 145600 | 85587 | 58158 | 483 | 64 |
| Available | 341280 | 184320 | 682560 | 744 | 3528 |
| Utilization | 43% | 47% | 8.6% | 65% | 2% |

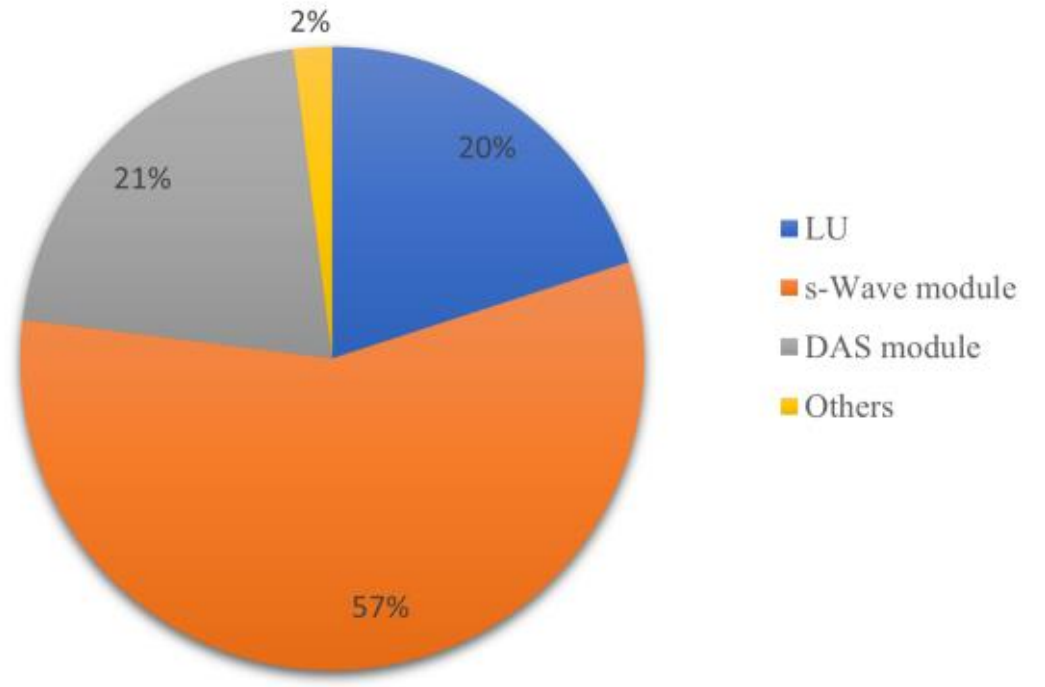

Fig. 12 Breakdown of the resource utilizations.

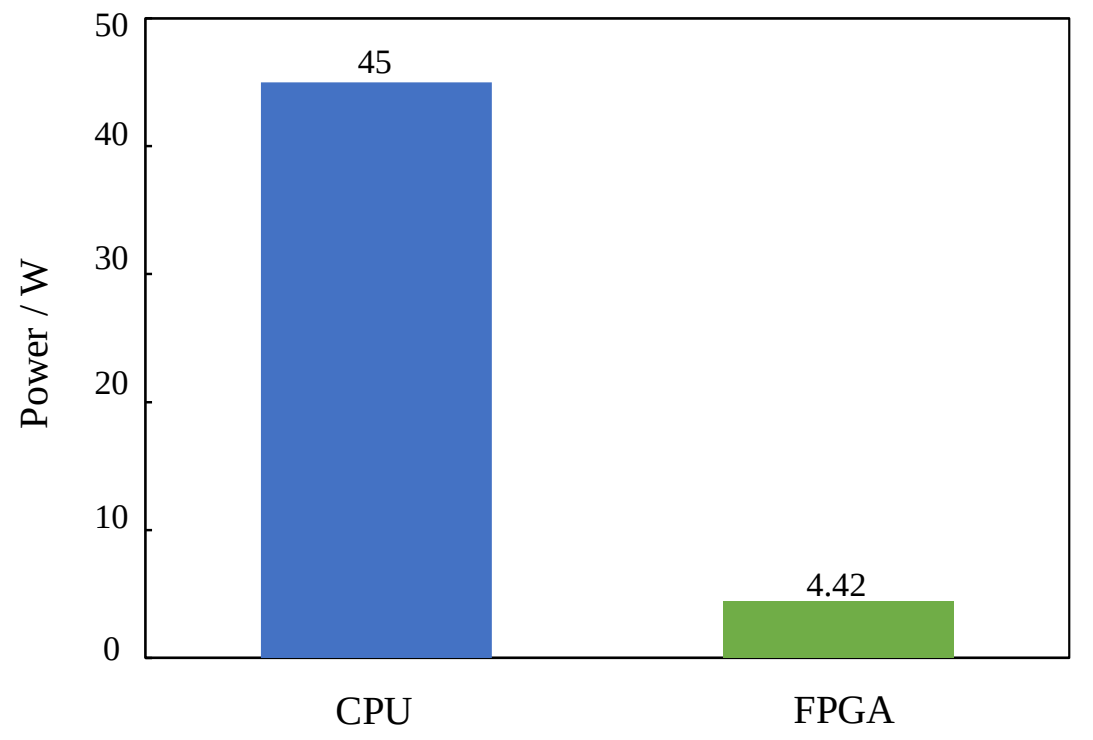

Fig. 13 Power comparison between CPU and FPGA.

## IV. CONCLUSION

In this paper, we present a pioneering implementation of a s-Wave-based iterative reconstruction algorithm for photoacoustic imaging on FPGA hardware, achieving remarkable advancements in both speed and energy efficiency compared to traditional CPU implementations. The proposed hardware architecture utilizes an innovative data reuse method to optimize the use of storage resources, making it possible to be implemented in on-chip memory. Moreover, leveraging the s-Wave simplification of the photoacoustic forward model allows the hardware implementation of the forward and backward processing of the imaging algorithm, along with other operations, to be efficiently executed using simple multiplication, addition, and memory access operations. The implementation results show that the proposed FPGA accelerated architecture is capable of achieving a maximum approximately 21 FPS, which underscores the feasibility of integrating the model-based algorithm into a palm-sized PAT system, paving the way for the development of real time, compact, high-quality PA imaging devices.

## ACKNOWLEDGMENT

This research was funded by National Natural Science Foundation of China (61805139), Shanghai Clinical Research and Trial Center (2022A0305-418-02), and Double First-Class Initiative Fund of ShanghaiTech University (2022X0203-904-04).